\documentclass[fleqn,twoside]{article}
\usepackage{espcrc2}

\usepackage{graphicx}
\usepackage[figuresright]{rotating}

\usepackage{amsmath,amssymb}

\title{$B_s$ mesons from the lattice: Excited states}

\author{J. Koponen\address[HY]{Department of Physical Sciences, \\ 
        P.O. Box 64, 00014 University of Helsinki, Finland}%
        \thanks{In collaboration with A. M. Green,
J. Ignatius, M. Jahma, C. McNeile, C. Michael and G. Thompson.
This work was supported by the Center for Scientific Computing,
Espoo, Finland, the Finnish Cultural Foundation, the Magnus Ehrnrooth
Foundation, the Academy of Finland (project 54038) and the EU
(grant HPRN-CT-2002-00311 EURIDICE).}, for UKQCD Collaboration
}

\begin{document}

%\begin{flushright}
%Preprint: HIP-2004-45/TH\\
%\end{flushright}
       
\begin{abstract}
The energies of different angular momentum states of a
heavy-light meson were measured on a lattice in \cite{MP}.
We have now continued this study using several different
lattices, quenched and unquenched, that have different physical
lattice sizes, clover coefficients, hopping parameters and quark--%
gluon couplings. The heavy quark is taken to be infinitely
heavy, whereas the light quark mass is approximately that 
of the strange quark. By interpolating in the heavy
and light quark masses we can thus compare the lattice results
with the $B_s$ meson. Most interesting is the lowest P-wave $B_s$
state, since it is possible that it lies below the $BK$ threshold
and hence is very narrow. Unfortunately, there are no experimental
results on P-wave $B_s$ mesons available at present.

In addition to the energy spectrum, we measured earlier also vector
(charge) and scalar (matter) radial distributions of the light quark
in the S-wave states of a heavy-light meson on a lattice \cite{GKMP}.
Now we are extending the study of radial distributions to P-wave states.
\vspace{1pc}
\end{abstract}

\maketitle

\section{Energies}

\begin{figure}
\centering
\caption{a) Two-point correlation function $C_2$;
b) Three-point correlation function $C_3$.}
\begin{tabular}{cc}
\includegraphics*[width=0.185\textwidth]{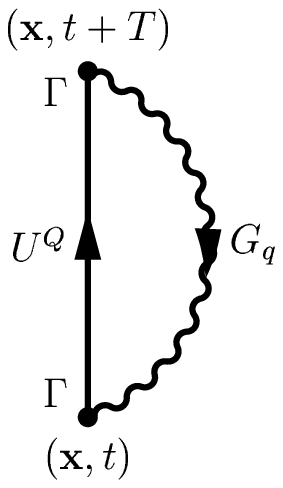}&
\includegraphics*[width=0.185\textwidth]{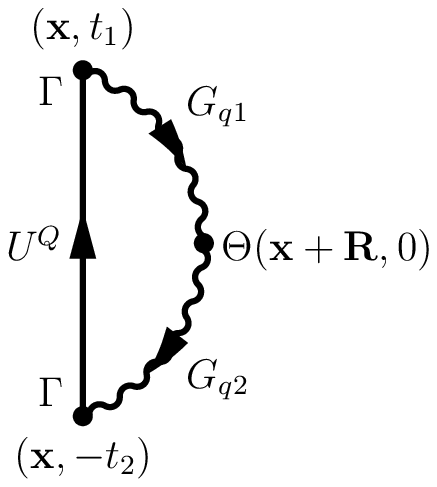}\\
 a) & b) \\ 
\end{tabular}
\label{fig_C2C3}
\vspace{-5mm}
\end{figure}

For simplicity we consider a system that consists of one infinitely heavy
quark (or anti-quark) and one light anti-quark (quark). The basic
quantity for evaluating the energies of this heavy-light system on a
lattice is the 2-point correlation function $C_2$ --- see
Fig.~\ref{fig_C2C3} a).
It is defined as
\begin{multline}
C_2(T)=\langle P_t\Gamma G_q(\mathbf{x},t+T,t)\\
\cdot P_{t+T}\Gamma^{\dag}U^Q(\mathbf{x},t,t+T)\rangle,
\end{multline}
where $U^Q$ is the heavy quark propagator and $G_q$ the light anti-quark
propagator. $P_t$ is a linear combination of products of gauge links at
time $t$ along paths $P$ and $\Gamma$ defines the spin structure of the
operator. The $\langle ...\rangle$ means the average over the whole lattice.
The energies are then extracted by fitting the $C_2$
with a sum of exponentials,
\begin{equation}
C_2(T)\approx\sum_{i=1}^{N_{\textrm{max}}}c_{i}\mathrm{e}^{-m_i T},
\label{equ_C2}
\end{equation}
where the number of exponentials, $N_{\textrm{max}}$, is 2, 3, or 4
and the time range $T\leq 10$.
$C_2(T)$ is in practice a $3\times 3$ matrix for different fuzzings
(the indices are omitted in Eq.~\ref{equ_C2} for clarity).
This fuzzing enables us to extract not only a more accurate value of
$m_1$, the energy of the ground state, but also an estimate of the
first radially excited state, $m_2$. After taking
the continuum limit we can interpolate in the light quark mass and use
charmed meson experimental results to interpolate in the heavy quark mass
to get a $B_s$ meson. The energy spectrum was measured using different
lattices: Q3 is a $16^3\times 24$ quenched lattice (for more details see
\cite{MP}), DF2 is a $16^3\times 24$ lattice and DF3, DF4 are $16^3\times 32$
unquenched lattices. DF refers to dynamical fermions (more
details can be found in \cite{GKMMT}). Parameters of the lattices are
given in Table~\ref{tbl_lattparam}.

\begin{table}
\centering
\caption{Lattice parameters. Here Q means quenched and
DF means unquenched (dynamical fermions).}
\begin{tabular}{cccc}
\hline
   &$\beta$&$C_{\textrm{SW}}$&$\kappa$\\
\hline
Q3 &5.7    &1.57             &0.14077 \\
DF2&5.2    &1.76             &0.1395  \\
DF3&5.2    &2.0171           &0.1350  \\
DF4&5.2    &2.0171           &0.1355  \\
\hline
\hline
   &$a$ [fm]&$m_q/m_s$&$r_0m_\pi$\\
\hline
Q3 &0.179(9)    &0.83     &1.555(6)\\
DF2&0.152(8)    &1.28     &1.94(3)\\
DF3&0.110(6)    &1.12     &1.93(3)\\
DF4&0.104(5)    &0.57     &1.48(3)\\
\hline
\end{tabular}
\label{tbl_lattparam}
\end{table}

\begin{figure}
\caption{The energy spectrum for different angular momentum states.
 The notation is explained in the text. 2S is the first radially
 excited $L=0$ state.
%Lattice parameters are given in Table~\ref{tbl_lattparam}.
 Here $r_0=0.525(25)$~fm.}
\centering
\includegraphics*[angle=-90,width=0.467\textwidth]{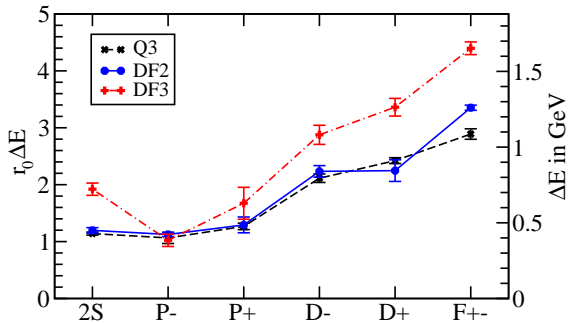}
\label{fig:energy}
\end{figure}

From some theoretical considerations \cite{Schnitzer}
it is expected that, for higher angular momentum states, the
multiplets should be inverted compared with the Coulomb spectrum.
If the potential is purely of the form $1/r$ the state L$+$ lies
always higher than L$-$. Here L$+$($-$) means that the light quark
spin couples to orbital angular momentum L giving the total $j=L\pm 1/2$.
(Since the heavy quark is infinitely heavy its spin does not play
a role here.)
However, in the case of QCD the confining potential is important for
large distances and thus L$-$ should eventually lie higher than L$+$ for
higher angular momentum states. Experimentally this inversion is not seen
for P-waves, and now the lattice measurements show that there is no
inversion in the D-wave states either. In fact, the D$+$ and D$-$
states seem to be nearly degenerate, \textit{i.e.} the spin-orbit
splitting is very small (see Fig.~\ref{fig:energy}). Note that the
spectrum in Fig.~\ref{fig:energy} shows an approximately linear rise
in excitation energy with L.

\begin{figure}
\caption{Interpolation from infinite mass to $m_Q=m_b$.
 The results at $m_Q=m_c$ are from charmed meson experiments.
 Note that the lowest $B_s$ P-wave state ($0^+$) lies below the $BK$
 threshold and should be very narrow.}
\centering
\includegraphics*[angle=-90,width=0.467\textwidth]{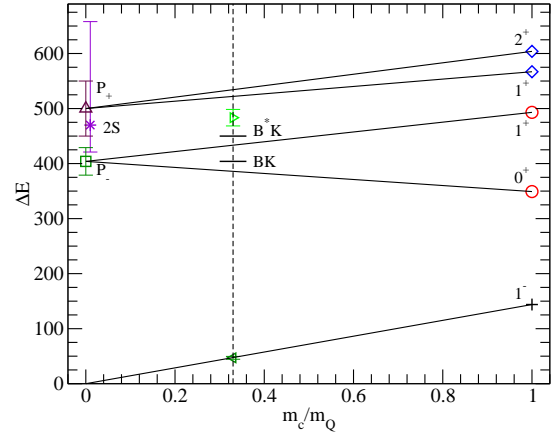}
\end{figure}

\section{Radial distributions}

\begin{figure}[t!]
\caption{Ground state vector and scalar radial distributions and discretized
 exponential fits.}
\begin{center}
\includegraphics*[angle=-90,width=0.465\textwidth]{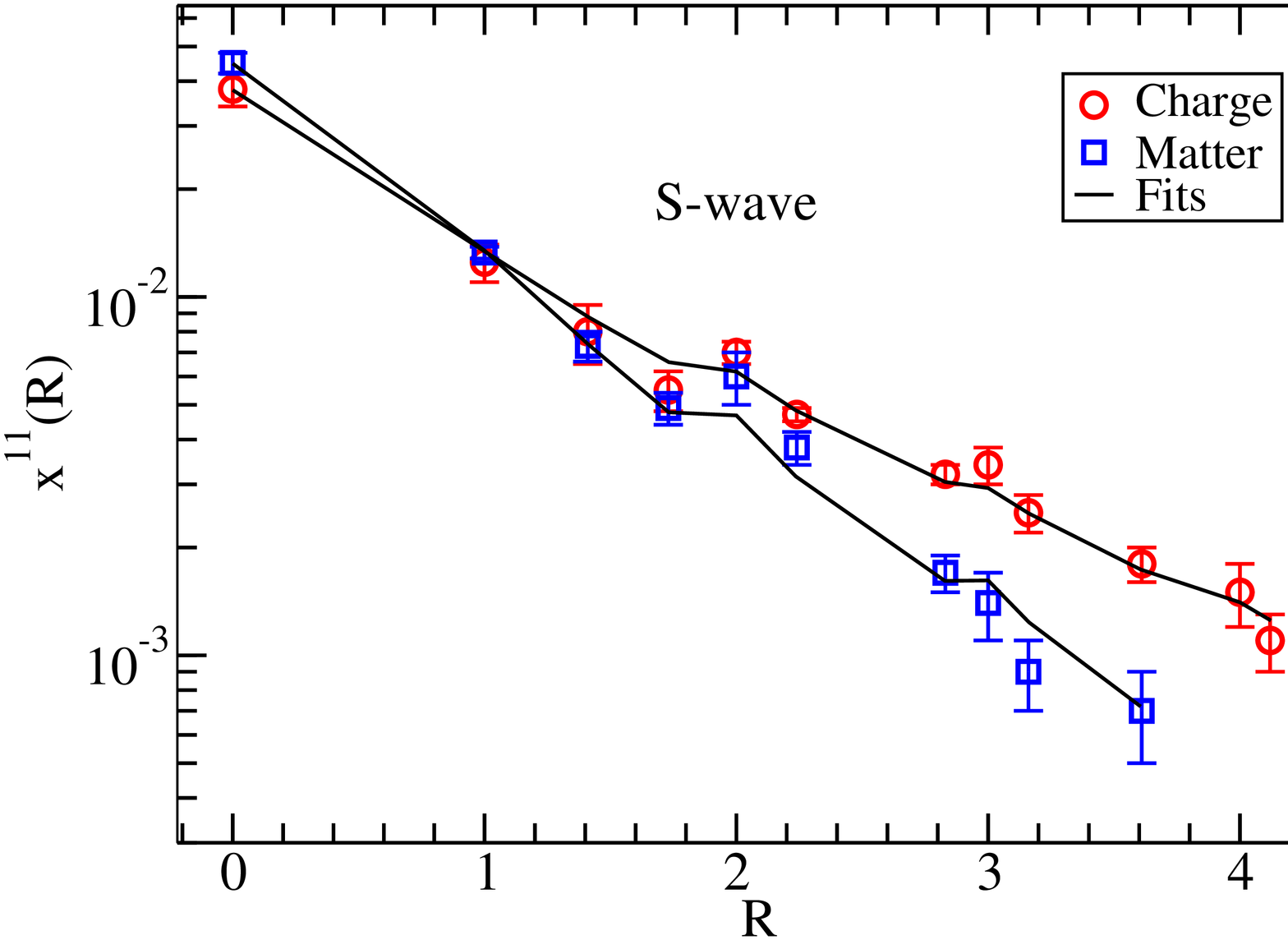}
\includegraphics*[angle=-90,width=0.465\textwidth]{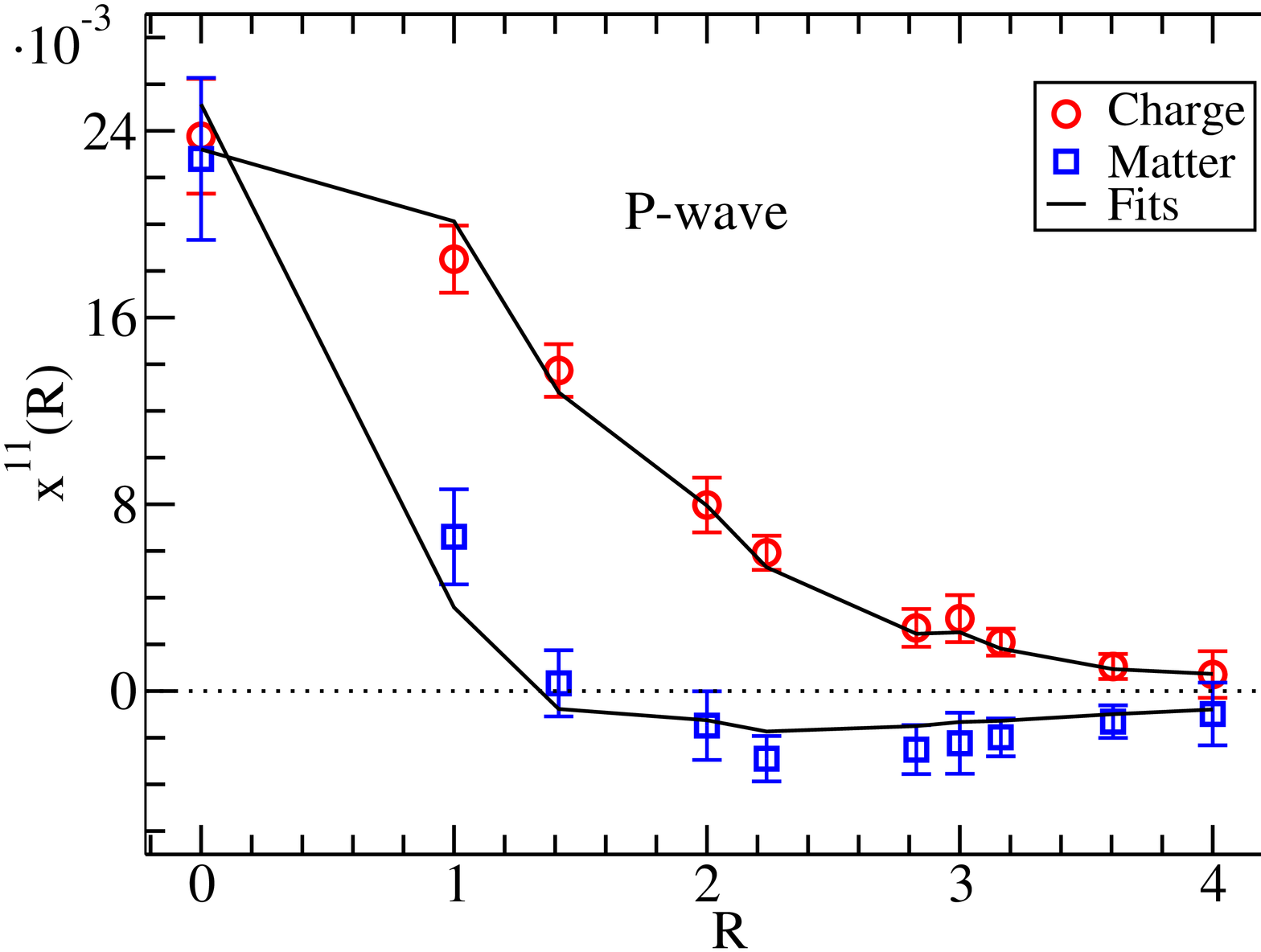}
\end{center}
\label{fig_X11}
\end{figure}

\begin{figure}[t!]
\caption{Radial distributions containing the first radially
 excited state (Ref.~\cite{GKMP} for the S-wave).}
\begin{center}
\includegraphics*[angle=-90,width=0.465\textwidth]{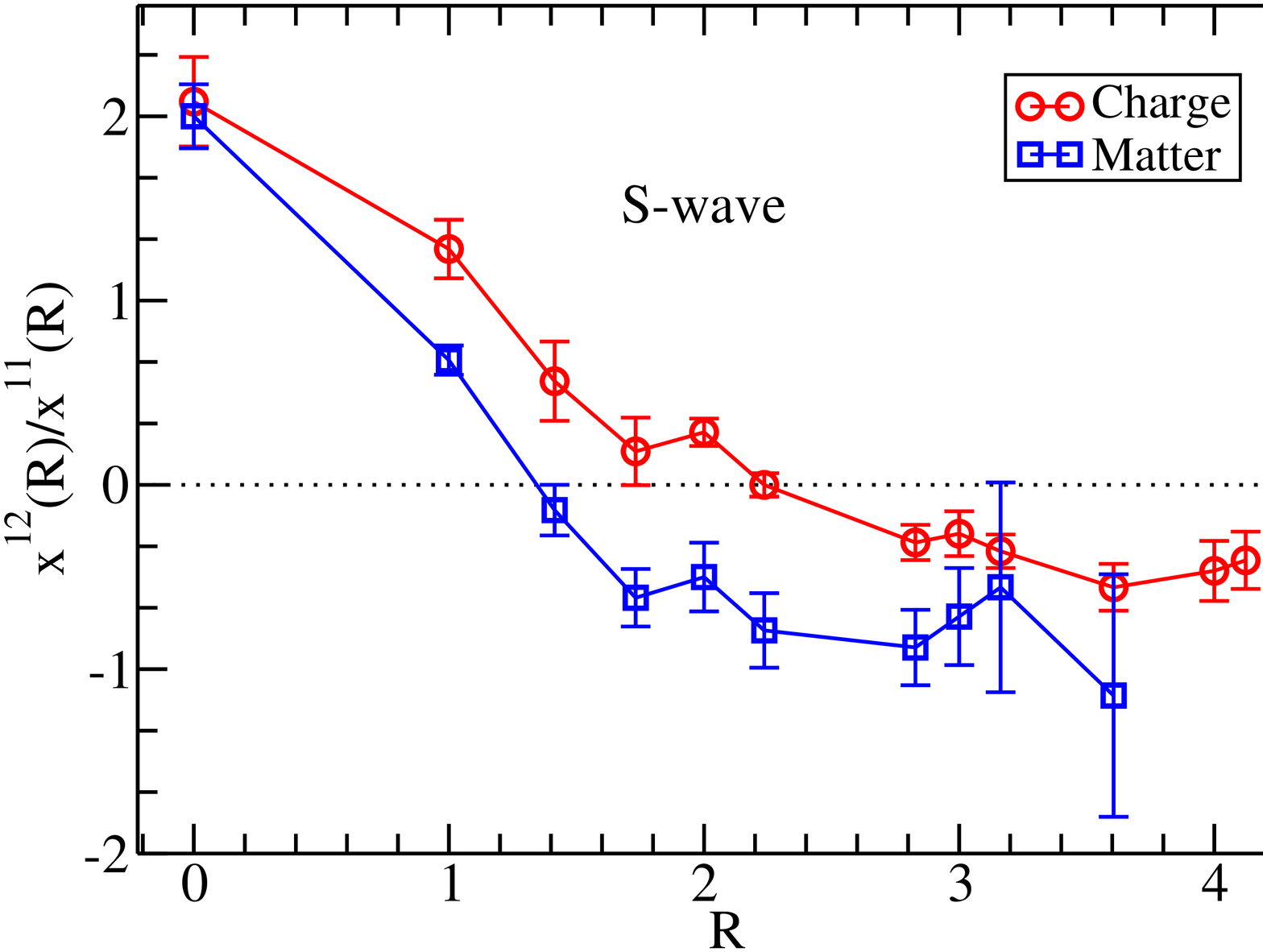}
\includegraphics*[angle=-90,width=0.465\textwidth]{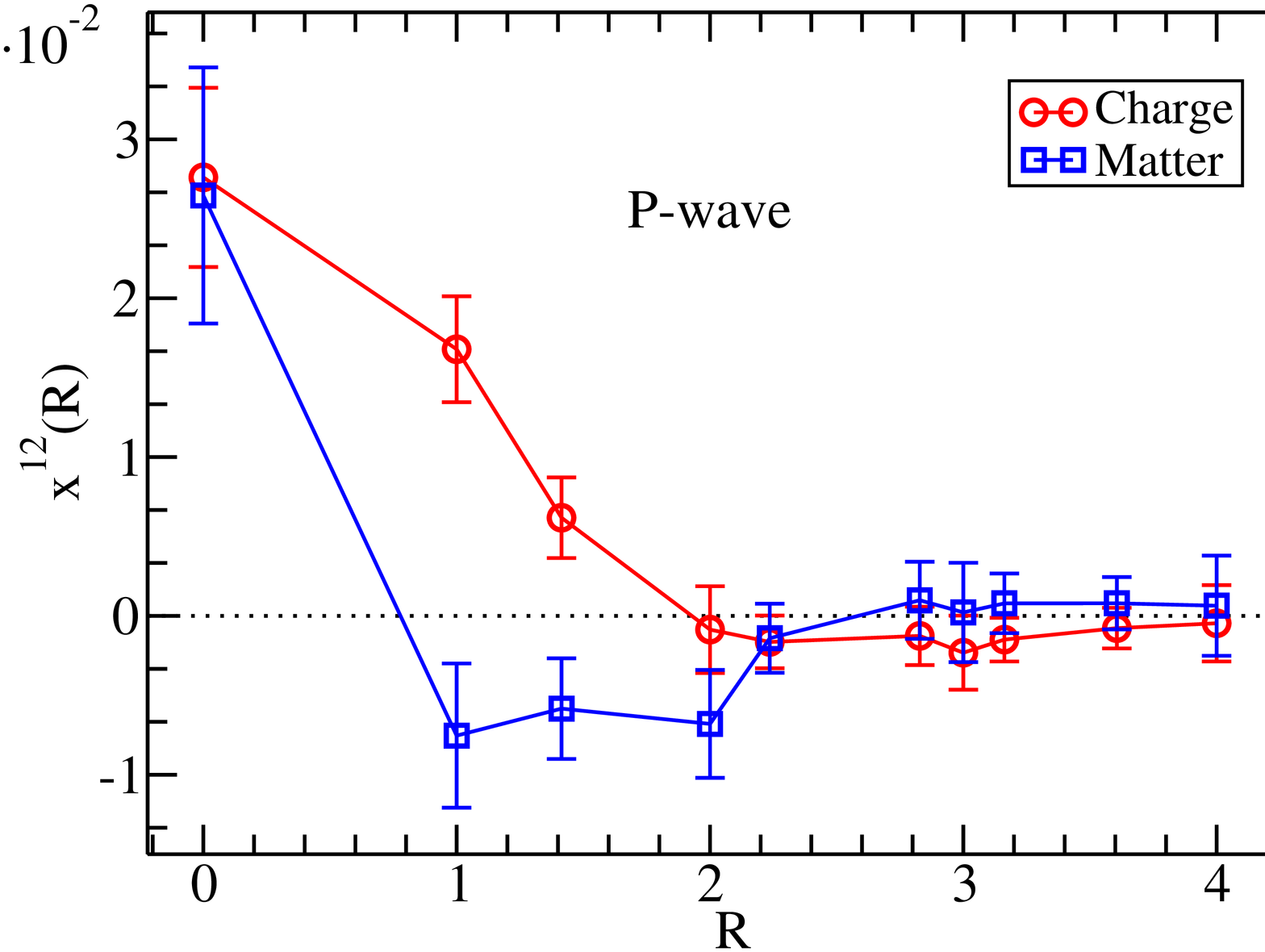}
\end{center}
\label{fig_X12}
\end{figure}

For evaluating the radial distributions of the light quark
a three-point correlation function is needed --- see
Fig.~\ref{fig_C2C3} b). It is defined as
\begin{equation}
C_3(R,T)=\langle \Gamma^{\dag}U^Q\Gamma G_{q1} \Theta
G_{q2}\rangle.
\end{equation}
We have now two light anti-quark propagators, $G_{q1}$ and $G_{q2}$,
and a probe $\Theta(R)$ at distance $R$ from the
static quark. We have used two probes:
\mbox{\boldmath{$\gamma$}}$_4$ for the vector (charge) and $\mathbf{1}$
for the scalar (matter) distribution. The radial distributions,
$x^{ij}(R)$'s, are then extracted by fitting the $C_3$ with
\begin{equation}
C_3(R,T)\approx\sum_{i,j=1}^{N_{\textrm{max}}}c_{i}\mathrm{e}^{-m_i t_1}%
x^{ij}(R)\mathrm{e}^{-m_j t_2}c_{j}
\end{equation}
--- see Figs.~\ref{fig_X11}, \ref{fig_X12}. The $m_i$'s and $c_i$'s
are from the best fit to $C_2$ (Eq.~\ref{equ_C2}). Here $x^{11}$ is
the ground state distribution and $x^{12}$, for example, is the
overlap between the ground state and the first excited state.

Fig.~\ref{fig_X11} compares the ground state vector (charge) distribution
with the scalar (matter) distribution. The matter distribution seems to
drop off faster than the charge distribution. However, at $R=0$ the charge
and matter distributions are roughly equal. Also note that the P-wave
matter distribution changes its sign. (Hence the logarithmic scale in
the S-wave plot but a linear scale in the P-wave plot in Fig.~\ref{fig_X11}.)
In Fig.~\ref{fig_X12} the charge and matter first excited state distributions 
are compared. As expected, a node is clearly seen in these excited state
distributions. The radial distributions shown here are for DF2, but the
analogous calculations for DF3 and DF4 are in progress. These distributions
can be now treated as ``experimental data'' that needs understanding.

\begin{figure}[t!]
\caption{Dirac equation numerical solutions, F and G, for a potential
$V=-e/R+b\cdot R$. Here $e=0.6\cdot\hslash c$, $b=(500~\textrm{MeV})^2$
and $m=100$~MeV.}
\begin{center}
\includegraphics*[width=0.37\textwidth]{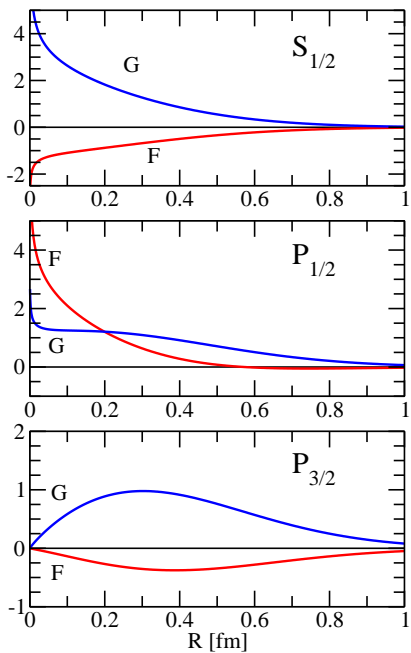}
\end{center}
\label{fig_Dirac2}
\end{figure}

\begin{figure}[t!]
\caption{Dirac equation numerical solutions for the same
potential as in Figure~\ref{fig_Dirac2} compared with lattice results.
$R$ is given in lattice units ($a \approx 0.15$~fm). Here the sign of
the S-wave matter distribution has been changed for clarity.}
\begin{center}
\includegraphics*[angle=-90,width=0.43\textwidth]{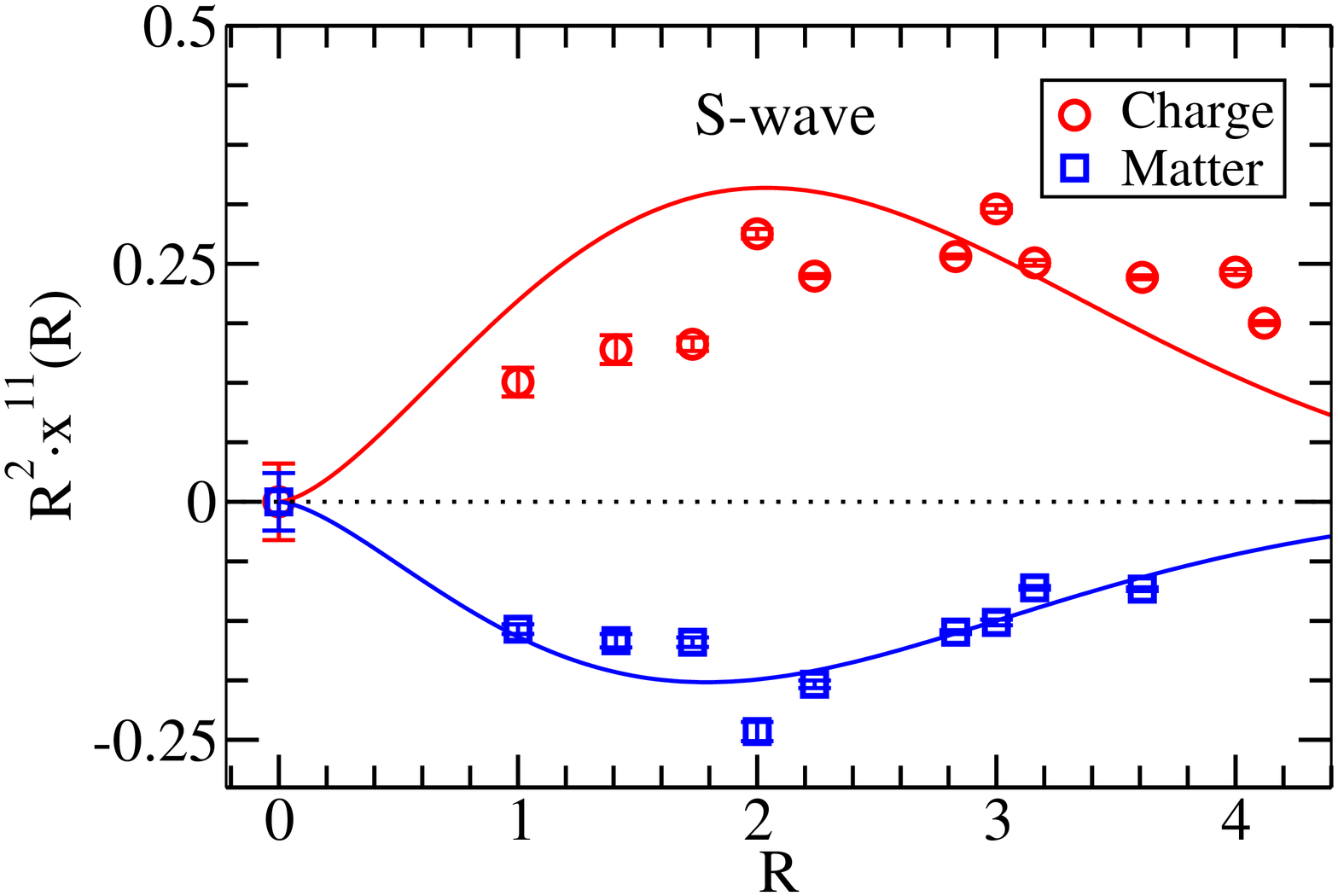}
\includegraphics*[angle=-90,width=0.43\textwidth]{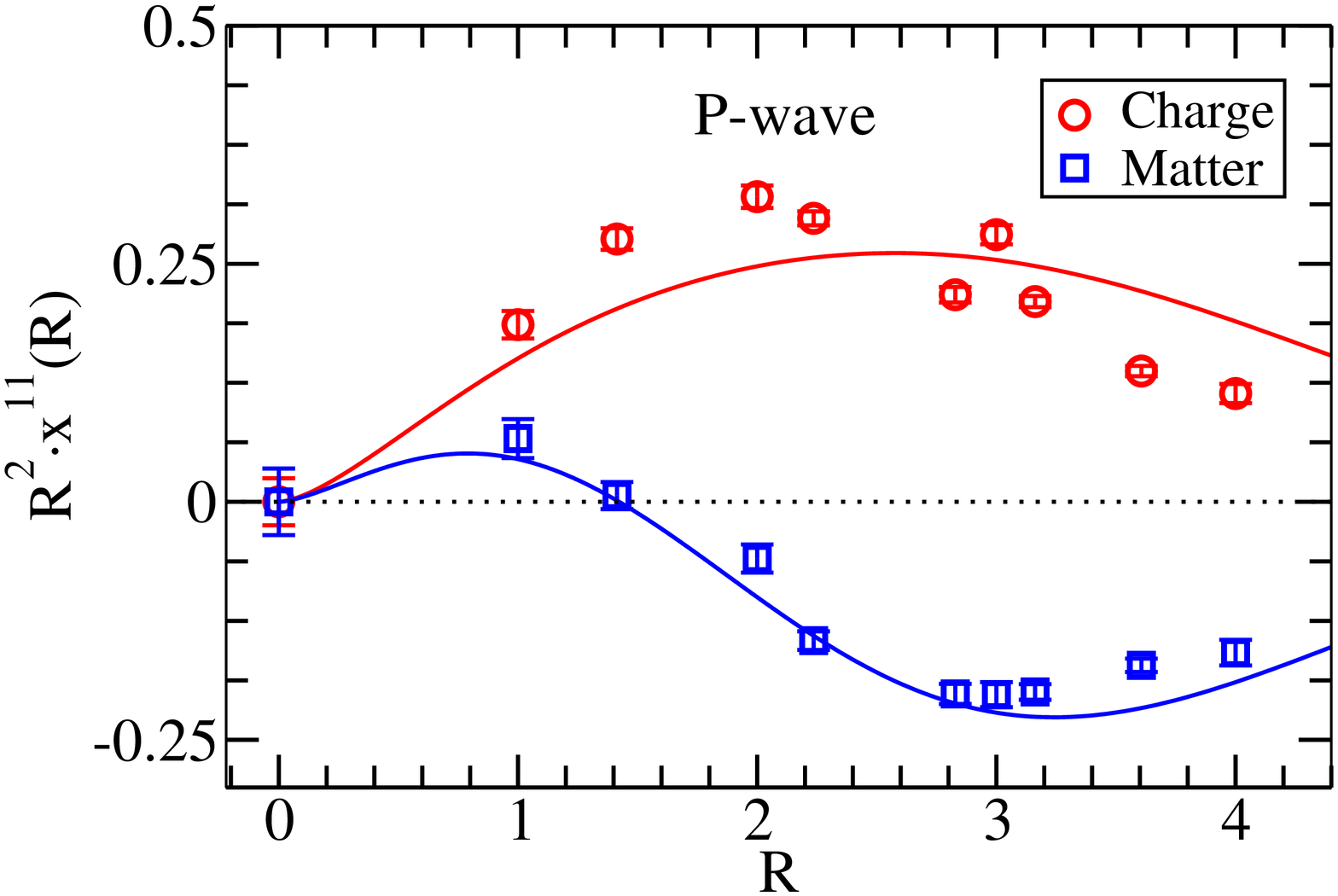}
\end{center}
\label{fig_Dirac}
\end{figure}

The lattice measurements, \textit{i.e.} the energy spectrum and the radial
distributions, can be used to test potential models. There are several
advantages for model makers: First of all, since the heavy quark is infinitely
heavy, we have essentially a one-body problem. Secondly, on the lattice
we know what we put in --- one heavy quark and one light anti-quark ---
which makes the lattice system much more simple than the real world.
We can, for example, try to use the Dirac equation to interpret the S- and
P-wave distributions.
If the solutions of the Dirac equation are called $G$ (the ``large''
component) and $F$ (the ``small'' component) we get
\begin{equation}
x^{11}(R)=G(R)^2+F(R)^2
\end{equation}
for the charge distribution and
\begin{equation}
x^{11}(R)=G(R)^2-F(R)^2
\end{equation}
for the matter distribution. These solutions are plotted in
Fig.~\ref{fig_Dirac2} for a potential $V=-e/R+b\cdot R$, where
$e=0.6\cdot\hslash c$, $b=(500~\textrm{MeV})^2$ and the mass
that appears in the Dirac equation is $m=100$~MeV.
The values of these parameters are not tuned yet, because at this
stage we only want to show that a reasonable set of parameters can
give a qualitative description of the lattice data. The qualitative
agreement is surprisingly good  --- see Fig.~\ref{fig_Dirac}.
The Dirac equation approach gives a natural explanation, for example,
to the fact that the matter distribution drops off faster than the
charge distribution. Also the change of sign in the P-wave matter
distribution comes out automatically.

\section{Main conclusions}

The key points are:
\begin{itemize}
\item
There should be several narrow $B_s$ states,
in particular the $0^+$ that lies below the $BK$ threshold.
\item
The spin-orbit splitting is very small, contrary to the
expectation predicted by a linearly rising potential that
is purely scalar.
\item
The radial distributions of S and P$-$ states can be qualitatively
understood by using a Dirac equation model.
\end{itemize}

The natural next step would be to calculate the radial distributions
for P$+$, D$-$, D$+$ and F-wave states.

\end{document}